\documentclass[a4paper,noshowpacs,showkeys,preprint,superscriptaddress,prb,nobibnotes]{revtex4}
\usepackage{graphicx}
\usepackage{amssymb}
\usepackage{amsmath}
\usepackage{epsfig}

\usepackage[usenames]{color}

\usepackage[top=1.5in, bottom=1.25in, left=1in, right=1in]{geometry}

\begin{document}

\title{Acoustic response in a a one-dimensional layered pseudo-Hermitian metamaterial containing defects}

\author{D.~Psiachos}
\email{dpsiachos@gmail.com}
\author{M.~M.~Sigalas}
\email{sigalas@upatras.gr}
\affiliation{Department of Materials Science, University of Patras, 26504, Rio, Greece}

\begin{abstract}
Using transfer-matrix methods, we investigate the response of a multilayered
metamaterial system containing defects to an incident acoustic plane wave at normal or oblique incidence. The transmission
response is composed of pass-bands with oscillatory behaviour, separated by
band gaps and covers a wide frequency range. The presence of gain and
loss in the layers leads to the emergence of symmetry breaking and re-entrant
phases. In the general case, a system containing defects will display a more general property, pseudo-Hermiticity (PH),
of which $\mathcal{PT}$ systems are a subset. In the PH-symmetric phase, unidirectional responses of 
the reflection, accomplished by reversing the parity $\mathcal{P}$, can be found but the response sometimes deviates from the predictions of simple scattering
theory which call for a pseudo-unitarity relation relating the transmission and the two directions of
reflections to hold. The converse of reversing the parity, reversing the time operator $\mathcal{T}$ in a spatially-asymmetric system within the PH-symmetric
regime can lead to different transmissions: a pass-band versus a stop-band. As regions
of stable PH-symmetric pass-band transmission oscillations occur over a wide spectral range, there
is a large flexibility in system parameters such as layer thicknesses, for leading to the desired unidirectional 
traits. In addition, we find that while defects in general lead to a near or complete loss of 
PH symmetry at all frequencies, they can be exploited to produce highly-sensitive responses, 
making such systems good candidates for sensor applications.
\end{abstract}

\maketitle

\section{Introduction}
Elastic wave propagation in layered media has parallels with wave propagation in 
a variety of physical systems:\textit{e.g.}
electrons in atomic lattices, electromagnetic waves in photonic lattices. The latter 
in particular has many parallels with elastic-wave propagation in phononic lattices,
as the general form of the field equations describing the wave propagation in the respective media
and also much of the formalism and methods of solution are common to both types
of systems, as are some of the proposed applications. 

Phononic lattices are particularly interesting as they may exhibit partial or total band gaps
in some energy regions, enabling the transmission of one type of polarization or none at all. Unlike
atomic crystals, homogeneous phononic materials do not exhibit total band gaps, necessitating
the construction of complex composites in order to access regimes where no transmission occurs.
The tunability of the characteristics of these highly-complex composite phononic materials results in behaviour 
completely unlike any material heretofore known, leading to these materials being termed 
``metamaterials". Elastic metamaterials exhibiting for example 
a negative effective mass density and bulk modulus have been constructed and some
proposed applications include systems such as superlenses~\cite{Croenne,Kaina} and the
redirection of sound: acoustic cloaking~\cite{ChenChanAppPh} or illusions~\cite{Layman}. Recently, the term
`metamaterials' has come to encompass a more general class of systems composed of materials, often
artificial, which combined lead to phenomena not found in the constituents themselves nor in any material 
found in nature~\cite{PopaNatureComm,CummerReview}. The correspondence 
with photonic systems is closer in the case of acoustic-wave propagation - in for example
a multilayer immersed in fluid - thereby enabling the equations of transformation acoustics to retain
their form under coordinate transformation~\cite{MiltonNJP}, a property particularly useful for the design of
acoustic cloaks.

Metamaterials may also be made ``active" leading to the production of gain or allowing absorption within them. 
However, because such materials have large amounts of losses owing to various types of material
imperfections, combining gain (G) and loss (L) materials constitutes
a approach to achieving reduced losses~\cite{TsironisPRL}. Parity-time ($\mathcal{PT}$) or the more general category
of pseudo-Hermitian (PH)~\cite{Deb} -symmetric         
systems, despite being non-Hermitian, can have real eigenvalues, as is necessary for 
propagation, as well as bound states. The $\mathcal{P}$ operator refers to reversing spatial coordinates, 
while the $\mathcal{T}$ operator performs complex
conjugation. Thus broken $\mathcal{P}$ in a system of multilayers with G/L can be induced by defects, while broken 
$\mathcal{T}$ can result from a reversal of the 
gain/loss elements. However, the combination of $\mathcal{P}$ and $\mathcal{T}$ may still be retained
even though one or both are broken separately. In a $\mathcal{PT}$ symmetric system the transfer matrix is
equal to its complex inverse~\cite{Mostaf} while the more general case, of a PH system,
is characterised by the transfer matrix not obeying the above property. For a general PH system, the Hamiltonian
obeys the property $H^\dagger=\eta H\eta^{-1}$ where $\eta$ is a Hermitian linear and invertible operator~\cite{Mostaf2002,Deb}, equal
to one when $H$ is equal to its Hermitian conjugate $H^\dagger$ while in $\mathcal{PT}$ symmetry, 
we have $(\mathcal{PT})H(\mathcal{PT})^{-1}=H.$ In the more general, PH symmetry, as in our system which displays
imperfect periodicity and/or localized defects, the operator $\eta$ is an arbitrary quantity. By modifying the system parameters, the system may
migrate from the $\mathcal{PT}$ or PH -symmetric phase into the broken phase and vice versa. The propagation in PH systems
differs from that in conventional ones in that it displays unidirectional properties, such as invisibility~\cite{Mostaf}. In addition, the total energy
is not conserved, oscillating about a mean value, owing to the non-orthogonality of the eigenmodes. A consequence
of the latter is an asymmetry in the propagation, known as non-reciprocal propagation~\cite{Ruter}. In the broken
phase, the propagation is unstable - greatly amplified or attenuated - and at these values of the wavevector the energy
dispersion relation takes on imaginary components. PH systems not displaying $\mathcal{PT}$ symmetry have recently started to be studied
in depth because of the greater flexibility~\cite{Trimer} afforded by removing the restrictions around the spatial arrangement of the
components and the presence of defects.

A recent
realization of acoustic metamaterials uses a dual microphone setup~\cite{Fleury}, one absorbing, the other
with gain, to achieve unidirectional propagation of 
sound. Unidirectional wave propagation in photonic gratings has been demonstrated
at the $\mathcal{PT}$ exceptional point~\cite{Regensburger, FengNatMater}, the transition point between
the $\mathcal{PT}$-symmetric and broken regime. Further, a $\mathcal{PT}$ system which 
on its own displays lasing, can become a coherent potential absorber~\cite{Longhi} when a second 
signal is injected in the reverse side and direction from the first signal, but
coherent with it. A system which behaves as a 
coherent potential absorber in the $\mathcal{PT}-$symmetric phase, has also been realized~\cite{Sun}. Some
related unidirectional applications in acoustics would include for example the suppression of 
echoes. Christensen \textit{et al}~\cite{ChristensenPRL} propose the use of electrically-biased piezoelectric semiconductors
in order to generate amplification or attenuation, demonstrating theoretically or through simulations 
that exceptional points and unidirectional
reflection-suppression may be achieved conveniently by this means. $\mathcal{PT}$-symmetric 
metamaterials have also been proposed for unidirectional 
acoustic cloaking applications~\cite{Zhu} or for the construction of superlenses,
capable of overcoming the diffraction limit with no losses~\cite{FleuryPRL}.

In this study, we investigate the propagation of acoustic waves through a one-dimensional multilayer system
including gain/loss (G/L) in order to establish the criteria for achieving unidirectional and/or orientation-dependent reflection and
transmission, for normal or oblique incidence. Previous theoretical investigations, albeit relying on simple
scattering models, based on two-port networks~\cite{Longhi,ChongPRL,LinPRL2011,Zhu,ZhaoPhysLett}, predict 
unidirectional reflection at the exceptional point~\cite{ChristensenPRL,ShiNatureComm}. Most
previous studies focus on balanced gain and loss, and find that a gap is necessary
between the gain and loss components firstly in order for $\mathcal{PT}$-symmetry breaking and reentrant phases 
to be present and secondly that the thickness of the spacing should be tuned in order to maximize the degree of directional
asymmetry~\cite{ShiNatureComm,ZhaoPhysLett}. By considering multilayers, we exploit the numerous
bands and wide ranges of pass bands in order to examine unidirectional behaviour and responses to
the presence of defects over multiple and wide-ranging
of $\mathcal{PT}$-symmetric regions. The defects or symmetry-breaking features we study consist of an asymmetry in the
reflection symmetry and in a modified thickness of some of the layers.

Many methods for determining the transmittance and reflectance properties of 
phononic systems exist~\cite{SigalasReview}: from multiple-scattering methods, to the finite-difference-time-domain (FDTD) method,
to transfer-matrix methods, etc. The transfer-matrix technique in particular has been applied to multilayered
systems in order to study elastic-wave propagation at normal and oblique incidence, where transverse modes
are activated, in
systems with defects or absorption~\cite{SigalasSoukoulis}.

\section{Theoretical Methods}
While there are many parallels with
optical waves, the motion of elastic waves is more complicated - requiring two
equations of motion to be satisfied for the time-harmonic field - even at normal
incidence, where both longitudinal and transverse modes within the elastic media are excited 
despite being independent.

\begin{figure}[htb]
\includegraphics[width=14cm]{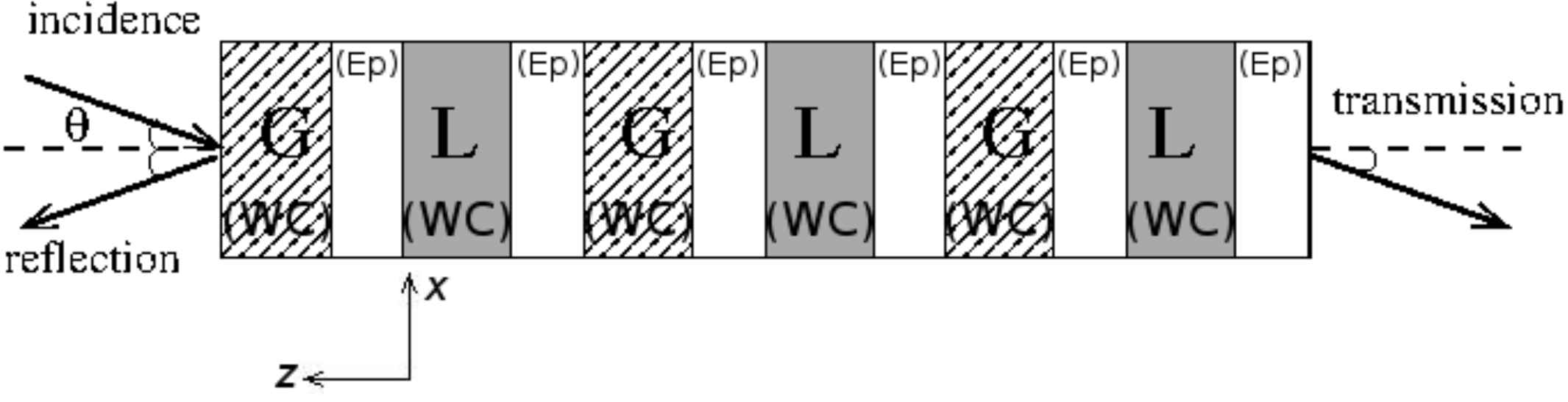}
\caption{Metamaterial system under study: impinging oblique acoustic wave onto a multilayer stack composed
of alternating gain (G) and loss (L) layers, separated by passive material. The materials are discussed in 
Sec.~\ref{Results}: WC (tungsten carbide) 
and Ep (epoxy thermoset).}
\label{fig1}
\end{figure}

In the following, we describe our implementation of the elastic-wave propagation.
We consider a system of $n-1$ layers with normal the $\hat{z}$ direction, extending along the negative $\hat{z}$ direction
as in Fig.~\ref{fig1}. Each layer is described by homogeneous mass density $\rho$ and
elastic properties: shear modulus $\mu$ and Lam\'{e} constant $\lambda$. The
longitudinal and transverse wave speeds are thus $c=\sqrt{(\lambda+2\mu)/\rho}$ and $b=\sqrt{\mu/\rho}$ respectively.  We further 
assume isotropic elasticity where the Lam\'{e} constant $\lambda=\frac{2\mu\nu}{1-2\nu}$ and the shear modulus
is $\mu=\frac{E}{2(1+\nu)}$ are expressed in terms of the Young's modulus $E$ and Poisson ratio $\nu$. A longitudinally-polarized 
plane wave of frequency $\omega$ is incident on the multilayer system from a liquid ambient medium
either normally or at an angle $\theta$ in the $xz$ plane and exits again in liquid. No other type of elastic wave can propagate
through liquids.
For such a polarization, the particle displacement
is in the $xz$ plane and there may be both shear and longitudinal 
modes present but only the latter in the liquid which terminates both ends of the layer. The 
displacement field may be split into
longitudinal $\phi$ and transverse $\vec{\psi}$ potentials
\begin{equation}
\vec{u}=\vec{\nabla}\phi+\vec{\nabla}\times\vec{\psi}
\label{u}
\end{equation}

and we set $\vec{\psi}=\psi \hat{y}$.
The wave equations for the potentials, assuming time-harmonic plane-waves are
\begin{align}\label{waveeq}
\nabla^2\phi+&k^2\phi=0,\;\hfill k\equiv \omega/c, \\ \nonumber
\nabla^2\psi+&\kappa^2\psi=0,\;\hfill \kappa\equiv \omega/b, 
\end{align}
and have the solutions
\begin{align}\label{prime} 
\phi_j&=\phi_j^\prime e^{i\alpha_j z}+\phi_j^{\prime\prime} e^{-i\alpha_j z},\; \hfill \alpha_j=\left(k_j^2-\xi^2\right)^{1/2},\;\hfill \xi=k_j\sin{\theta}\\ \nonumber
\psi_j&=\psi_j^\prime e^{i\beta_j z}+\psi_j^{\prime\prime} e^{-i\beta_j z},\; \hfill \beta_j=\left(\kappa_j^2-\chi^2\right)^{1/2},\;\hfill \chi=\kappa_j\sin{\theta}
\end{align}
for each layer $j$, including the terminating ambient media $j=1$ and $j=n+1,$ where the primed quantities are amplitudes. For example,
for a longitudinal wave incident at the liquid half-space $n+1$, $r=\phi_{n+1}^{\prime}$ and $t=\phi_{1}^{\prime\prime}$
are the expressions for the longitudinal-mode reflection and transmission coefficients in an outgoing liquid layer $j=1$.

Through transformations from the $\{\phi^\prime,\phi^{\prime\prime},\psi^\prime,\psi^{\prime\prime}\}$ basis, 
a transfer matrix for the passage of a wave through one, and then by repeated application, through
 the whole system of $n-1$ layers may be constructed in
terms of the displacements~Eq.\ref{u}
and the stresses
\begin{align}\label{Z1}
Z_x&=\mu\left(\partial u_x/\partial z+\partial u_z/\partial x\right) \hfill \qquad \mathrm {and}\\ \nonumber
Z_z&=\lambda\left(\partial u_x/\partial x+\partial u_z/\partial z\right)+2\mu\partial u_z/\partial z
\end{align}
as 
\begin{equation}
\begin{pmatrix}
u_x^{(n)}\\
u_z^{(n)}\\
Z_z^{(n)}\\
Z_x^{(n)}
\end{pmatrix}=\underline{\underline{A}}
\begin{pmatrix}
u_x^{(1)}\\
u_z^{(1)}\\
Z_z^{(1)}\\
Z_x^{(1)}
\end{pmatrix}
\label{Mmatrix2}
\end{equation}
where $\underline{\underline{A}}$ is the transfer matrix through the entire multilayer
(see ch.1, sec.8 in Ref.\onlinecite{Brekhovskikh}). In the system depicted in Fig.~\ref{fig1} the transfer
matrix is not equal to its complex inverse, and this is why we will refer to it as ``PH symmetric" rather 
than ``$\mathcal{PT}$ symmetric". Otherwise, the transfer matrix of a PH-symmetric system displays the same key property as 
in a ``$\mathcal{PT}$ symmetric" system - the system displays regions of stable propagation (symmetric phase) when the eigenvalues
of the transfer matrix are unimodular, separated by regions
of unstable propagation (broken phase) when the eigenvalues deviate from this condition. 

The boundary conditions applied are the continuity of the displacements $u_x$ and $u_z,$ and that of the stresses
$Z_x,$ $Z_z$ across
the $(n,n+1)$ boundary. In the case of a liquid ambient medium, only the latter two are required, with
$Z_x$ additionally being zero as shear stresses are not supported in the liquid. When this
is done, we arrive at expressions for the reflection and transmission from the incident and outgoing
sides respectively of the multilayer system.

Thus, the analytical approach to calculating the response of a multilayered stack
embedded in fluid to an incident plane 
acoustic wave may be summarized as follows: Beginning from the
potentials satisfying the two-dimensional wave equations Eq.\ref{waveeq}, 
we construct the state variables - displacement (Eq.~\ref{u}) and stress (Eq.~\ref{Z1}). The coefficients of the state
variables then comprise the transfer matrix in a given layer. By repeated multiplication, the transfer matrix
through the entire multilayer system is constructed. Then, 
subject to the boundary conditions
at the edges of the stack according to which the transverse stresses
are zero at the fluid-stack interfaces, and the other components are continuous, the transmission and reflection coefficients
may be determined by inverting the system of equations for the stress and displacement to obtain the amplitudes
in Eq.~\ref{prime} which comprise them, as noted above. The formalism 
is well-documented, for example in Refs.~\onlinecite{Brekhovskikh,Munjal}. The formalism is fully general so as to
work also with solid ambient media and to
transverse incident waves. In the general case, there are interconversions between 
transverse and longitudinal reflections and transmissions which will be interesting to study.

\section{Results}
\label{Results}
The response of multilayer system composed of alternating layers with the elastic properties of tungsten carbide (WC) and
epoxy thermoset (Ep) respectively to an impingent acoustic wave was examined for various
incidence angles. The schematic is as in Fig.~\ref{fig1}. These particular elastic properties (rather than the materials themselves, which may
well end up being artificial in nature in some future implementation) were chosen in order to investigate a typical phononic system,
namely with drastically different elastic properties in order to achieve
the requisite opening up of spectral gaps~\cite{SigalasReview}. The parameters of these materials are given in 
Table~\ref{parameters}. The layer thicknesses have been all set to the same value unless otherwise noted. The frequencies
at which the responses occur scale inversely with the value of the thickness $l$ used (\textit{viz.} units
on graphs' axes). When
including G/L, through the addition of an imaginary component of positive (G) or negative (L) sign 
to the Young's modulus, we 
work in the region of balanced gain/loss, by imposing alternating G/L on the WC layers. While loss can be imparted
using dissipation, gain requires the use of active elements. The amount of G/L imparted here
is in line with that achievable for the tunable effective bulk modulus in a recent implementation of a metamaterial~\cite{PopaPRB}.
\begin{table}
\begin{tabular}{l|| l| l| l|l|l}
Material& $E$ (GPa)& Im($E$)& $\nu$ & $\rho$ (Mg/m$^3$)& layer thickness\\
\hline
WC (tungsten carbide)&550&$\pm$30&0.21&15.5&$l$\\
Ep (epoxy thermoset) &3.5&0&0.25&1.2&$l$
\end{tabular}
\caption{Parameters of the WC/Ep multilayer system. $E$ (Im$E$) is the real (imaginary - when activated) part 
of the Young's modulus, $\nu$ the Poisson Ratio, $\rho$ the mass density, $l$ the layer thickness (uniform in this study
unless otherwise noted). The parameters
of the fluid (water) at both ends are $c=$1.5 m/s, $\rho=$1 Mg/m$^3$.}
\label{parameters}
\end{table}

\begin{figure}[htb]
\includegraphics[width=14cm]{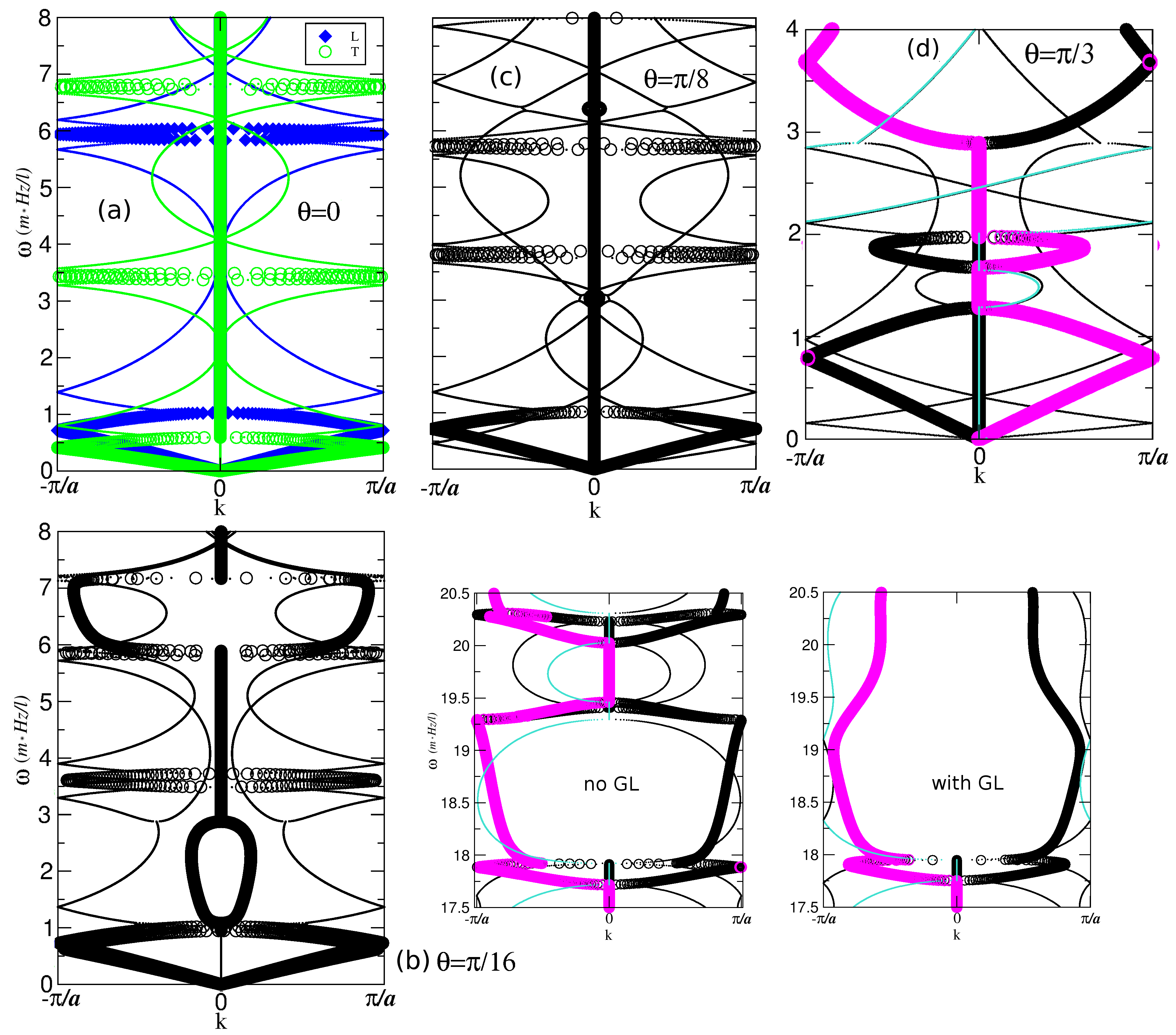}
\caption{Dispersion relation $k(\omega)$ (albeit with inverted axes) of a WC$^-$/Ep/WC$^+$/Ep infinite system for different incidence
angles. Real (imaginary) parts of $k$ are displayed with thick (thin) symbols. In (a), we use a different colour scheme in order to highlight the
longitudinal (L) and transverse (T) modes. (b) is comprised of three panels. In (b) and (d) we superimpose as pink (real) and turquoise (imaginary) colours the components of one of the
four solutions for $k$. The lattice
constant $a$ is the thickness of the tetralayer.}
\label{fig2}
\end{figure}

For small
incidence angles we find regions of $\mathcal{PT}$ or more generally PH-symmetry separated by 
broken symmetry.  The characterization of $\mathcal{PT}$ or PH-symmetry was done
in two ways. In Fig.~\ref{fig2} we show the dispersion relation $k(\omega)$, albeit with inverted axes for
simpler interpretation, for $\theta=0$, $\theta=\pi/16,$ $\theta=\pi/8,$
and $\theta=\pi/3$ for systems including G/L and in the first two cases we find regions of purely real wavevectors $k$, which is essential for the propagation
of these modes with no amplification or attenuation. The dispersion relation was calculated using the Floquet
theorem, in an approach modified from Ref.~\onlinecite{Esquivel}, \textit{viz.}
\begin{equation}
\begin{pmatrix}
u_x^{(n)}\\
u_z^{(n)}\\
Z_z^{(n)}\\
Z_x^{(n)}
\end{pmatrix}=\exp{(ikL)}
\begin{pmatrix}
u_x^{(n-1)}\\
u_z^{(n-1)}\\
Z_z^{(n-1)}\\
Z_x^{(n-1)}
\end{pmatrix}
\label{Floquet}
\end{equation}

for $L$ the total thickness of one multilayer period. By using the analog of the 
transfer-matrix $\underline{\underline{A}}$ but for one period, called $\underline{\underline{a}},$
the solutions for the wavevector $k$ are found to satisfy the equation
\begin{equation}
\mathrm{det}[\underline{\underline{a}}-\underline{\underline{I}}\exp{(ikL)}]=0
\label{det}
\end{equation}
 where $\underline{\underline{I}}$
is the identity matrix.

In the same ranges that all four wavevector solutions were purely real, 
the eigenvalues of the transfer matrix were all
simultaneously unimodular, ensuring stable propagation up to infinite stack length. Without G/L the band structures
remained to a large extent exactly the same, although for some higher $\omega$ values, deviations can be seen, as in Fig.~\ref{fig2}b. When
G/L is absent however, the systems display none of the characteristics of PH systems,
only those of conventional materials.

\begin{figure}[htb]
\includegraphics[width=8cm]{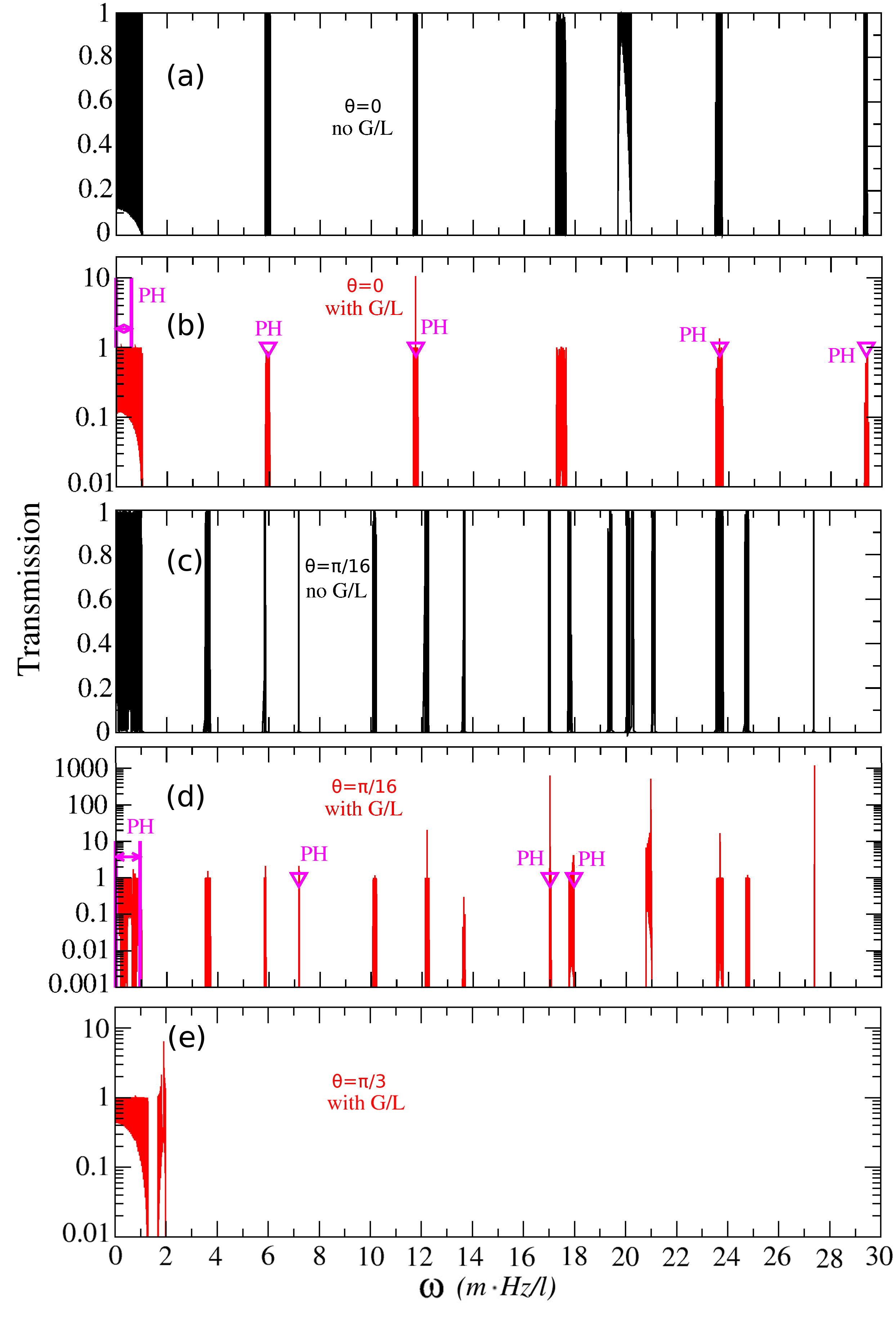}
\caption{Transmission amplitude $T$ for 15 tetralayers WC$^-$/Ep/WC$^+$/Ep for (a)-(b), $\theta=0$ with and
without gain/loss (G/L), (c)-(d), $\theta=\pi/16.$ In (e) we show the system for $\theta=\pi/3$ for G/L included. Regions 
of PH-symmetry corresponding to the infinite system
are delineated by the pink bars and/or the triangles for when the region is very narrow (the entire band is PH symmetric).}
\label{fig3}
\end{figure}

 In Fig.\ref{fig2}a ($\theta=0$), we show the longitudinal (L) and transverse (T) modes
separately. The imaginary part of the transverse mode is non-zero beginning at $\omega=0.593.$ However,
as seen in the calculation of the transmission response, albeit for a finite system, in Fig.~\ref{fig3}a-b, 
oscillations in the pass-band still occur in the region up until $\omega=1.027,$ the point
at which the imaginary
part of the longitudinal mode becomes nonzero since the transverse mode is not excited. For
$\theta=\pi/16,$ there is full PH symmetry up until $\omega=0.948$ (Fig.~\ref{fig2}b), reduced from the `effective' 
1.027 in the $\theta=0$ case
due to the mixing of the L and T modes. In Fig.~\ref{fig3}c-d we
see that a greater number of bands appear for $\theta=\pi/16$ compared with the $\theta=0$ case. Comparing with the cases
of no gain/loss, we find that the locations of the bands are mostly unaltered in Fig.~\ref{fig3}. However, in some cases,
some of the pass-bands disappear upon going from the passive to the G/L system. In Fig.~\ref{fig3}c there are bands \textit{e.g.}
at $19.29< \omega< 19.46$ and $20.02<\omega< 20.31,$ amongst others, which are not present
when G/L are present (Fig.~\ref{fig3}d). Referring to the band structure in Fig.~\ref{fig2}b, where this frequency range is 
displayed in the second and third panels, we see that in these ranges the real part of the wavevector consists of 
two nearly coincident bands with small curvature in each of these ranges only in the system with no G/L. In these figures,
one out of the four complex wavevector solutions is highlighted as an example. In the
case of no G/L, the real part of each wavevector does not overlap with its imaginary part and the mode propagates. In contrast,
with G/L included, such a solution does not exist and the mode is not propagating. In the cases shown in Fig.~\ref{fig3}b,d
which are not classified as PH-symmetric, each wavevector solution is complex, but propagates because the real and imaginary parts of the same
solution do not overlap as explained above. 

In Fig.~\ref{fig3}, regions classified as PH-symmetric
often display transmission greater than one. However, the material displacements and stresses $(u_x,u_z,Z_z,Z_x)$ 
remain finite for arbitrary multilayer repetition, in contrast with regions not classified as PH-symmetric, where
these parameters blow up.
  For large incidence angles, such as in Fig.~\ref{fig2}c for $\theta=\pi/8,$ and Fig.~\ref{fig2}d for $\theta=\pi/3$ 
we do not find
any regions of real wavevectors; even at the low $\omega$ region, the imaginary parts have moved
off the $k=0$ axis and any propagating solutions as a result of nonoverlapping real and imaginary
parts of the solutions for $k(\omega)$ (see Fig.\ref{fig2}d and ~\ref{fig3}e) are not PH-symmetric and the material parameters
blow up.

\begin{figure}[htb]
\includegraphics[width=12cm]{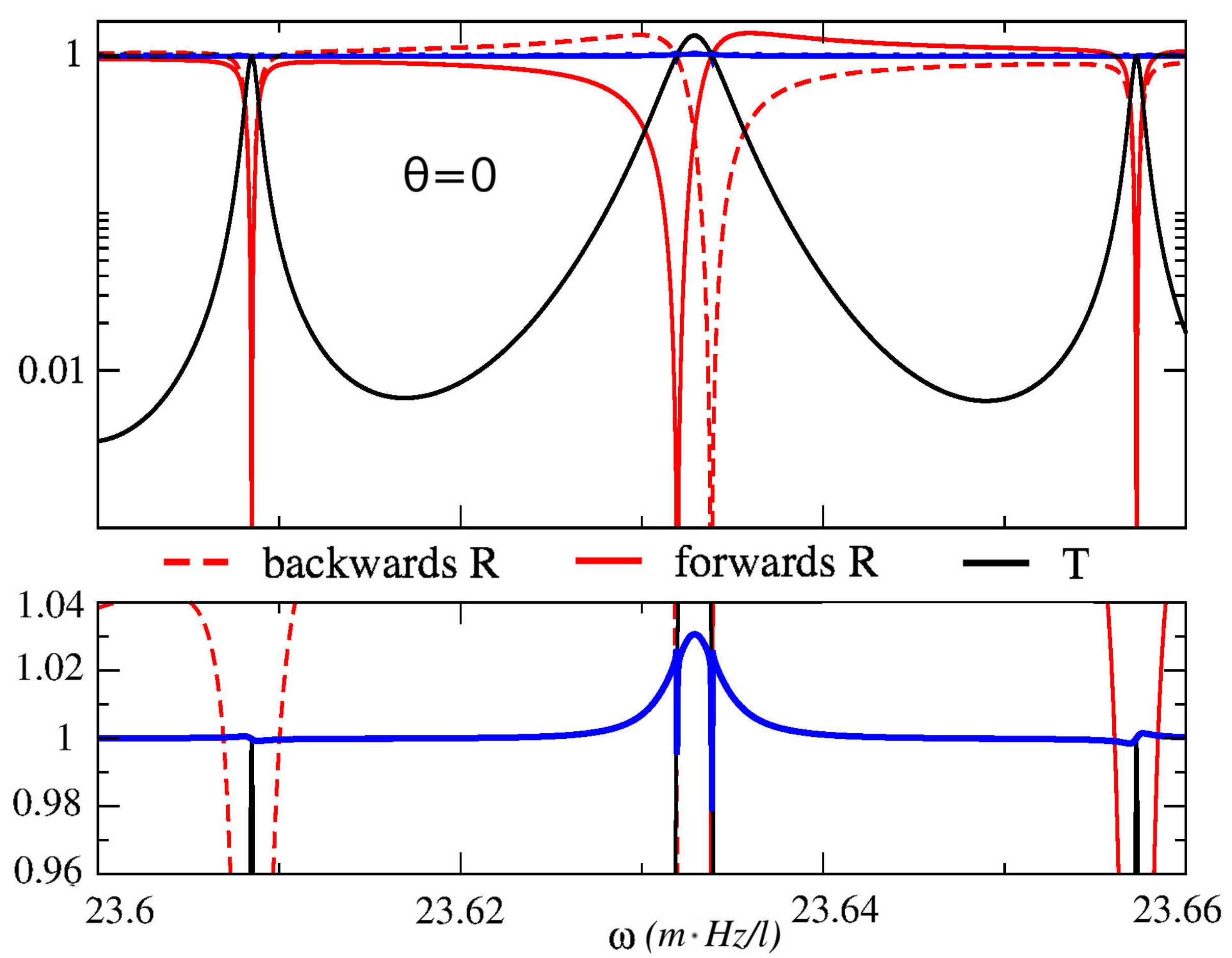}
\caption{Reflection and transmission amplitudes for 15 tetralayers WC$^-$/Ep/WC$^+$/Ep for $\theta=0,$
whether approached from the gain side for WC, \textit{i.e.} Ep/WC$^+$\dots (backwards) or loss \textit{i.e.} WC$^-$/Ep\dots 
(forwards) direction. In this example the system is in the PH-symmetric regime. The thick blue
curve is the expression $T\pm\sqrt{R_F R_B}$ according to whether $T<1$ (plus) or $T>1$ (minus). The figure on the
bottom is a close-up of the top figure.}
\label{fig4}
\end{figure}

Signature features of systems with balanced gain and loss - whether in the broken or the PH-symmetric
phase - include reflection ($R=|r|^2$) and transmission ($T=|t|^2$) amplitudes achieving
values greater than one and the reflection being direction-dependent: R from the right side bring unequal
to that from the left side. We did not find a directionality occuring for the T, consistent with what
the theory from models of $\mathcal{PT}$-symmetric scattering, based on two-port networks,
predicts~\cite{Longhi,ChongPRL,LinPRL2011,Zhu,ZhaoPhysLett}. However, the same theory
predicts that the product of the R is zero when $T=1,$ which we find to usually, but not always
be the case in a PH system. More generally, the above models predict that T and the forwards and backwards reflections $R_F,$ 
$R_B$ are related by the pseudo-unitarity relation $|T-1|=\sqrt{R_F R_B}.$ However, the two-port scattering
formalism and the aforementioned pseudo-unitary relation hold only for idealized $\mathcal{PT}$ systems, not for general
PH systems which arise through the inclusion of
defects such as in Ref.~\onlinecite{GeFeng}. This is entirely due to the fact that the forward and backward 
transfer matrices do not obey the required relations in such systems. In Fig.~\ref{fig4} we show how the pseudo-unitarity relation is
broken near strong T or R resonances, something which holds in the PH-symmetric regime (shown) as well as in the
broken regime, as well as for different incidence angles and other frequency ranges.

\begin{figure}[htb]
\includegraphics[width=13cm]{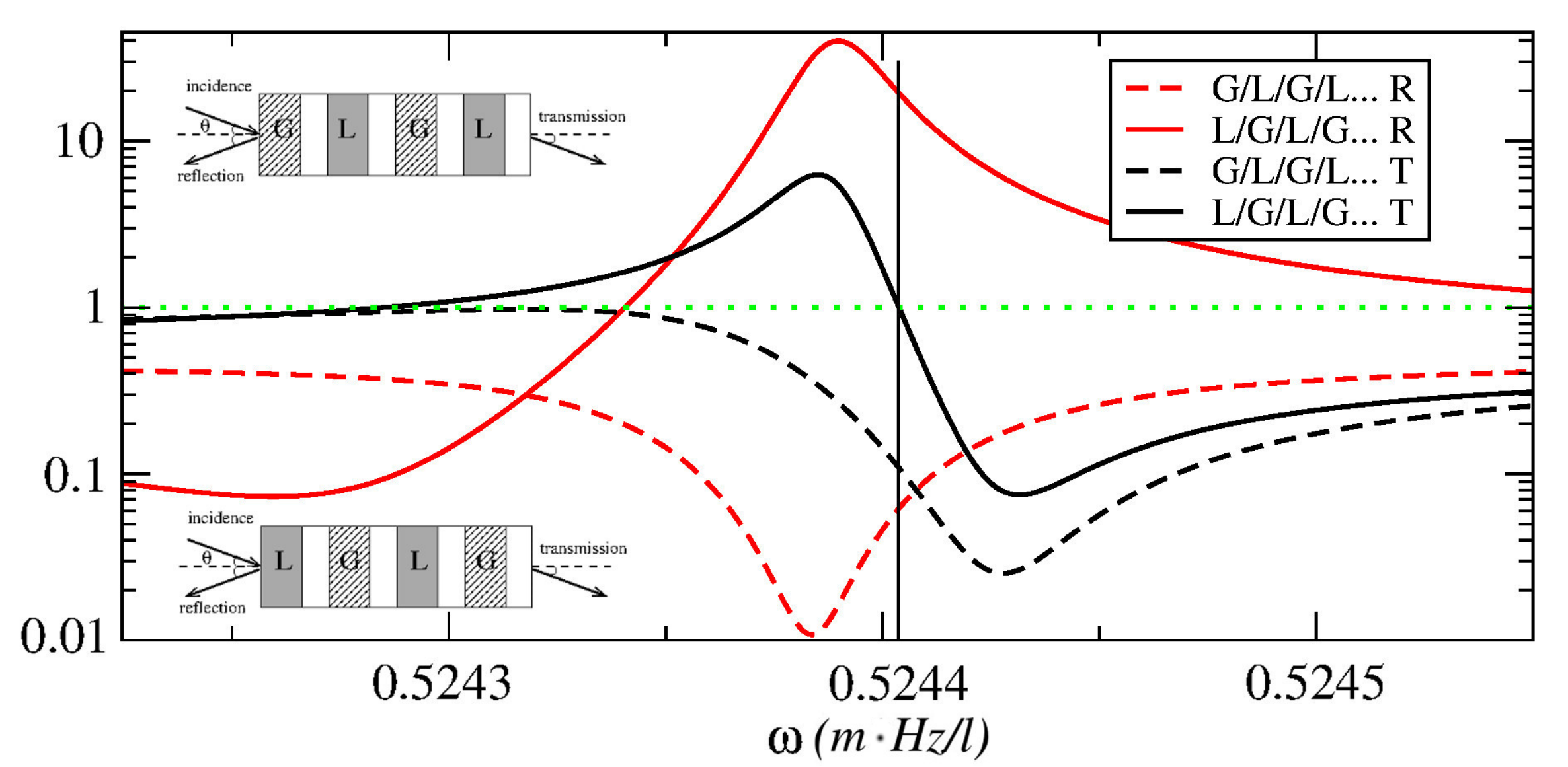}
\caption{Reflection and transmission amplitudes within a PH-symmetric region 
for 15 tetralayers WC/Ep/WC/Ep... for $\theta=\pi/16$ 
whether the ordering of the WC begins with gain or with loss. The two different orientations
are as depicted as an inset. The vertical line 
corresponds to one of the T being equal to 1. The green dotted line is a guide to the eye.}
\label{fig5}
\end{figure}

Additionally we find that the inclusion of defects in general reduces the range of or completely destroys the 
PH symmetry. This is evidenced by the eigenvalues of the transfer matrix no longer being unimodular at all,
or being so in a reduced range. Defects we have studied include having a layer with  modified 
thickness. We focus here mainly on systems in which we calculate the response with respect to a reversal
of spatial coordinates \textit{i.e.} $\mathcal{P}.$ However, in some special systems not containing defects \textit{per se} but 
having pre-existing $\mathcal{P}$ asymmetry, such as an even
number of layers WC/Ep, reversing the $\mathcal{T}$ (index of refraction, or G/L ordering here), has the effect of
breaking the symmetry in not only R but T as well. We indeed find
a dramatic effect of orientation: in which one G/L ordering is close to or greater than one while the other is zero for both R and T. Such
a system might be implemented with piezo-electric induced imaginary components - demonstrated to be able to tune
the G/L character~\cite{ChristensenPRL}. We further find 
that dramatic differences in the two responses is only possible for $T$ when $\theta\ne 0$; at normal incidence the two T differ but only
very slightly. Only the R have dramatic differences at normal incidence. In Fig.~\ref{fig5}
we show the R and T for $\theta=\pi/16$ where at the point indicated, one ordering has a very large R (19.26) and T=1 and
the other has low R and T (0.06 and 0.1 respectively).

We find, in general, that this interesting behaviour, where the response is 
either ``unidirectional" or specific to the G/L ordering, 
doesn't occur just at a single point as in prior works~\cite{ChristensenPRL,Zhu} but in many places, leading
to a flexible manipulation of resonances. Other resonances are not all as sharp as the one in Fig.~\ref{fig5} and it
would be also interesting to investigate the resonances produced by different choices of material parameters. The 
spacing of the Ep layer can be tuned so that the desired unidirectional
behaviour occurs at other frequencies - in contrast with a single gain / loss component, the 
multilayer system can access many such candidate frequencies owing to the presence of multiple pass bands
and the tuning of the spacing is not critically important.
\begin{figure}[htb]
\includegraphics[width=13cm]{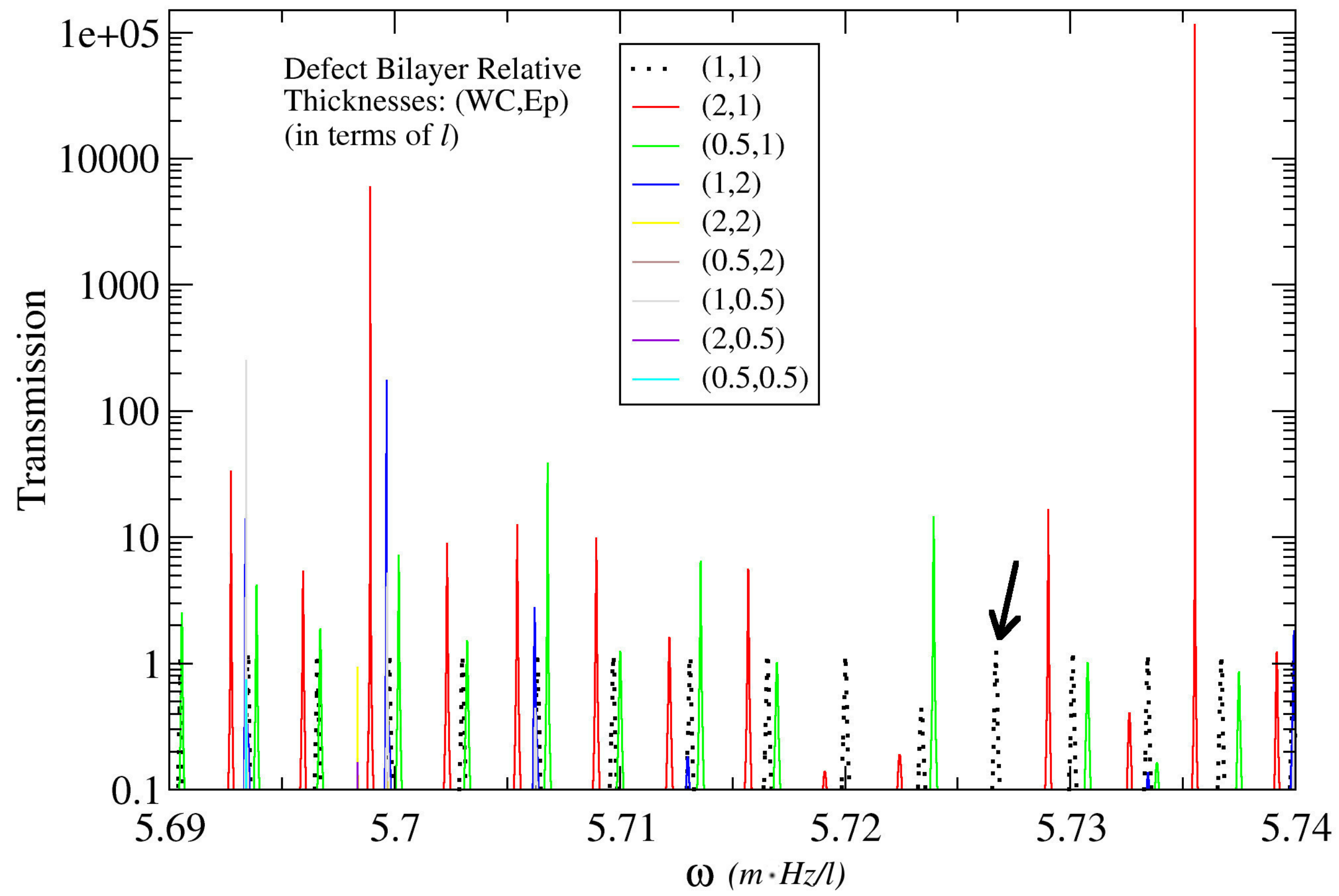}
\caption{Transmission amplitudes $T$ with defects. Region shown is near the highest resonance (1.23 at $\omega$=5.7267, marked with arrow) 
of the undefected system (black dotted line). 
The system consists of 25 tetralayers WC+/Ep/WC-/Ep... and $\theta=\pi/8.$ The defect consists of different relative thicknesses
of a (WC,Ep) bilayer near the central region (49th-50th layer), compared with the undefected value of (WC,Ep)=(1,1) - units in terms of $l$.}
\label{fig6}
\end{figure} 

Defects of thickness of the components were also examined. Out of the entire 25-tetralayer (WC+/Ep/WC-/Ep...) system
one WC-Ep bilayer was chosen to have variable thickness, either half or double the thickness of the 
rest of the layers. Although the system is in the broken phase owing
to its sensitivity to defects,
giant increases in transmission resonances, sometimes of several orders of magnitude greater than 
the uniform system, were observed. In Fig.~\ref{fig6} we show that for an incidence angle of $\pi/8,$ a 
(WC,Ep)=(2,1) defect bilayer thickness - henceforth all thickness units are in terms of $l$ - 
can lead to an enormous resonance in the $\omega$=5.65-5.8 band of the undefected system. This 
frequency range was studied because it is transmissive at all incidence 
angles studied (see Fig.~\ref{fig3} but also for $\theta=\pi/8$ and $\pi/4$). Specifically, a doubling
of the WC thickness in one of the central bilayers led to an enormous resonance in the transmission of 115417, versus
1.23 the highest point in the undefected system in this range. Similar results were found for other incidence angles,
including $\theta=0.$ However, we 
found that there is a very large variation with position - the largest resonances for when the thickness
defect was located in neighbouring bilayers varied by orders of magnitude and sometimes corresponded to different
values of the defect thicknesses. There was a variation amongst angles as well: for $\theta=0$ the preferred defect
thickness for achieving the highest resonance was (WC,Ep)=(2,2) while for $\pi/16$ and $\pi/8$ it was fairly
inconsistent but often (0.5,2) was preferred for $\pi/16$ and (0.5,1) for $\pi/8.$ For $\theta=\pi/4$ however, the
preferred thickness was most usually (0.5,1). This means that this particular defect-layer thickness responds to transverse modes better
than the (2,2) combination. For all angles and generally for all defect locations, the thickness of 0.5 for the Ep defect layer was found
to lead to the smallest resonances.

\begin{figure}[htb]
\includegraphics[width=10cm]{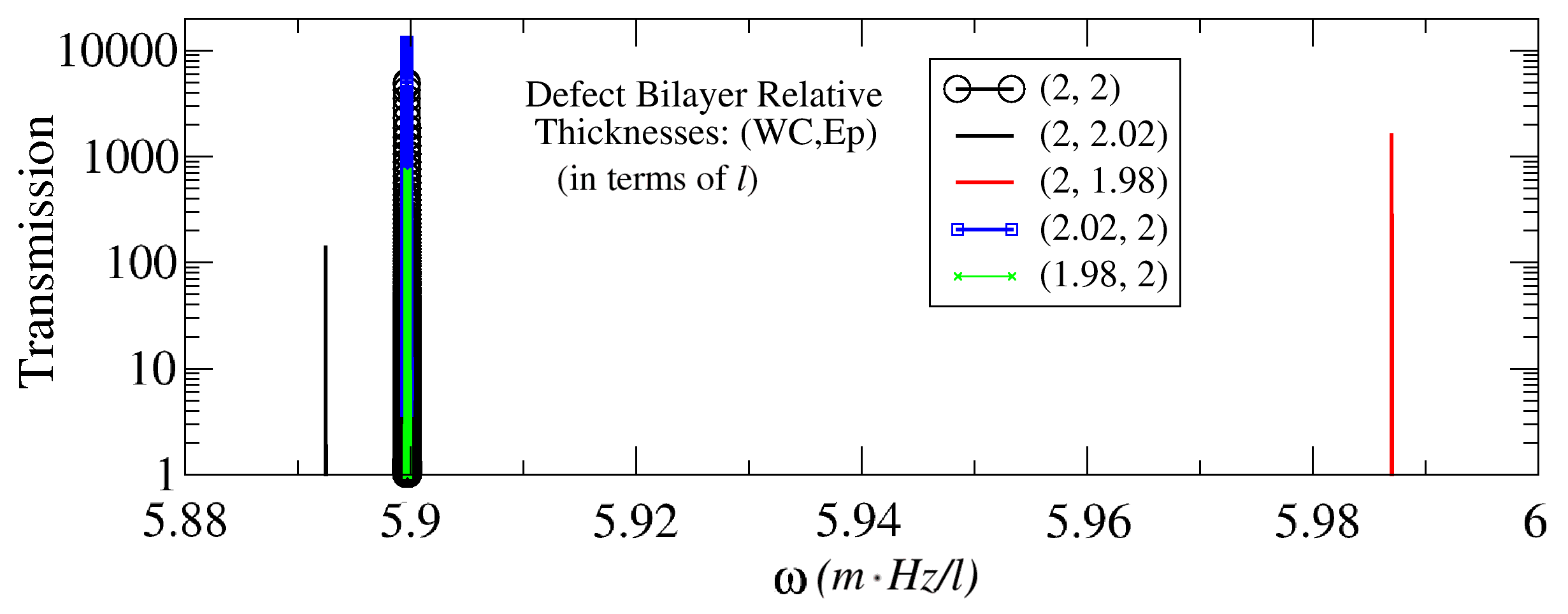}
\caption{Resonances of the transmission amplitude $T$ for one percent changes in the defect-layer thicknesses, about the optimal
thickness (WC,Ep)=(2,2) for the 55th-56th layers for a system of 
25 tetralayers WC+/Ep/WC-/Ep... for $\theta=0.$ The three curves with the same Ep thickness lie on top
of each other. Shown are only the highest resonances for each case.}
\label{fig7}
\end{figure} 
In recent work, Zhao \textit{et al}~\cite{ZhaoPhysLett}
have shown that small defect-layer thickness variations can dramatically alter the left/right reflectance ratio. We 
have examined how sensitive the giant transmission resonances are to small (one percent) variations in the thickness. In Fig.~\ref{fig7}
we examine a one percent variation in the defect thickness which led to the largest resonance for the 55th-56th (WC-Ep) layers for $\theta=0,$
\textit{i.e.} (WC,Ep)=(2,2). As shown in the figure, the response is highly sensitive to changes in thickness. In addition, for
changes in thickness of Ep, the resonance frequency displays a shift, while for changes in WC thickness with fixed Ep thickness it does not. 
These
giant resonances in the transmission and general sensitivity to defects can be exploited to fabricate ultra-sensitive
biochemical sensors~\cite{Sigalas-bio1,Sigalas-bio2}. Another interesting application of the present system would be for 
ultrasensitive touch-screens, since very small variations in thickness would result in extreme changes in transmission or reflection.

\section{Conclusions}
We calculate the reflection and transmission response to obliquely-incident acoustic waves 
in a pseudo-Hermitian (PH) system of multilayered gain/loss components separated by a passive spacer by solving
the elastodynamic equations in combination with the transfer-matrix technique. We find transmission oscillations
occuring in bands, covering a wide frequency range and in many cases the propagation is stable (PH-symmetric
phase) while in other cases it is unstable (broken phase) leading to a blowing-up of the material parameters. We 
also note some deviations from the simple scattering models, namely that the pseudo-unitary condition relating
the transmission to the reflections in the two spatial propagation directions does not exactly hold, particularly
near strong transmission or reflection resonances. On the other hand, for fixed propagation direction but reversed
time, we find a breaking of the reflection and transmission symmetry, 
sometimes simultaneously and at one ordering severely suppressed. As our system is multilayered in nature,
it can access a wide range of frequency bands, thus calibrating
the thickness of the passive spacer is not crucial for its operation, neither is its
operation at a single exceptional point. The presence of defects as well as their location within the system was 
found to have a profound effect on the transmission 
response: changing the 
thickness of one passive layer shifts transmission resonances to different frequencies, while even
very small changes in thickness were found to produce great sensitivity in the responses.

\end{document}